\documentclass[
superscriptaddress,
prl,aps,footinbib,twocolumn
]{revtex4}

\usepackage{graphicx}
\usepackage{dcolumn}
\usepackage{bm}
\usepackage{color}
\usepackage{amsthm,amsmath,amssymb}
\usepackage{mathrsfs}

\begin{document}

\preprint{APS/123-QED}

\title{Quantum Induced Coherence Light Detection and Ranging}

\affiliation{Interdisciplinary Center for Quantum Information, State Key Laboratory of Modern Optical Instrumentation, and Zhejiang Province Key Laboratory of Quantum Technology and Device, School of Physics, Zhejiang University, Hangzhou 310027, Zhejiang Province, China \\}
\affiliation{ZJU-Hangzhou Global Science and Technology Innovation Center, College of Information Science and Electronic Engineering,
Zhejiang University, Hangzhou 310027, China \\}
\affiliation{Texas A$\&$M University, 4242 TAMU, College Station, Texas 77840, USA \\}
\affiliation{Hefei National Laboratory, Hefei 230088, China \\}
\affiliation{CAS Center for Excellence in Topological Quantum Computation, University of Chinese Academy of Sciences, Beijing 100190, China}

\author{Gewei Qian$^1$, Xingqi Xu$^{1,*}$, Shun-An Zhu$^1$, Chenran Xu$^1$, Fei Gao$^2$, V.~V.~Yakovlev$^3$, Xu Liu$^1$, Shi-Yao Zhu$^{1,4}$ and Da-Wei Wang$^{1,4,5,\dagger}$}

\date{\today}

\begin{abstract}
Quantum illumination has been proposed and demonstrated to improve the signal-to-noise ratio (SNR) in light detection and ranging (LiDAR). When relying on coincidence detection, such a quantum LiDAR is limited by the response time of the detector and suffers from jamming noise. Inspired by the Zou-Wang-Mandel experiment, we design, construct and validate a quantum induced coherence (QuIC) LiDAR which is inherently immune to ambient and jamming noises. {In traditional LiDAR the direct detection of the reflected probe photons suffers from deteriorating SNR for increasing background noise. In QuIC LiDAR we circumvent this obstacle by only detecting the entangled reference photons, whose single-photon interference fringes are used to obtain the distance of the object, while the reflected probe photons are used to erase path information of the reference photons. In consequence, the noise accompanying the reflected probe light has no effect on the detected signal. We demonstrate such noise resilience with both LED and laser light to mimic the background noise and jamming attack.} The proposed method paves a new way of battling noise in precise quantum electromagnetic sensing and ranging.

\end{abstract}

\maketitle

\begin{figure*}[htbp]
\begin{center}
\includegraphics[scale=1,angle=0,width=0.8\textwidth]{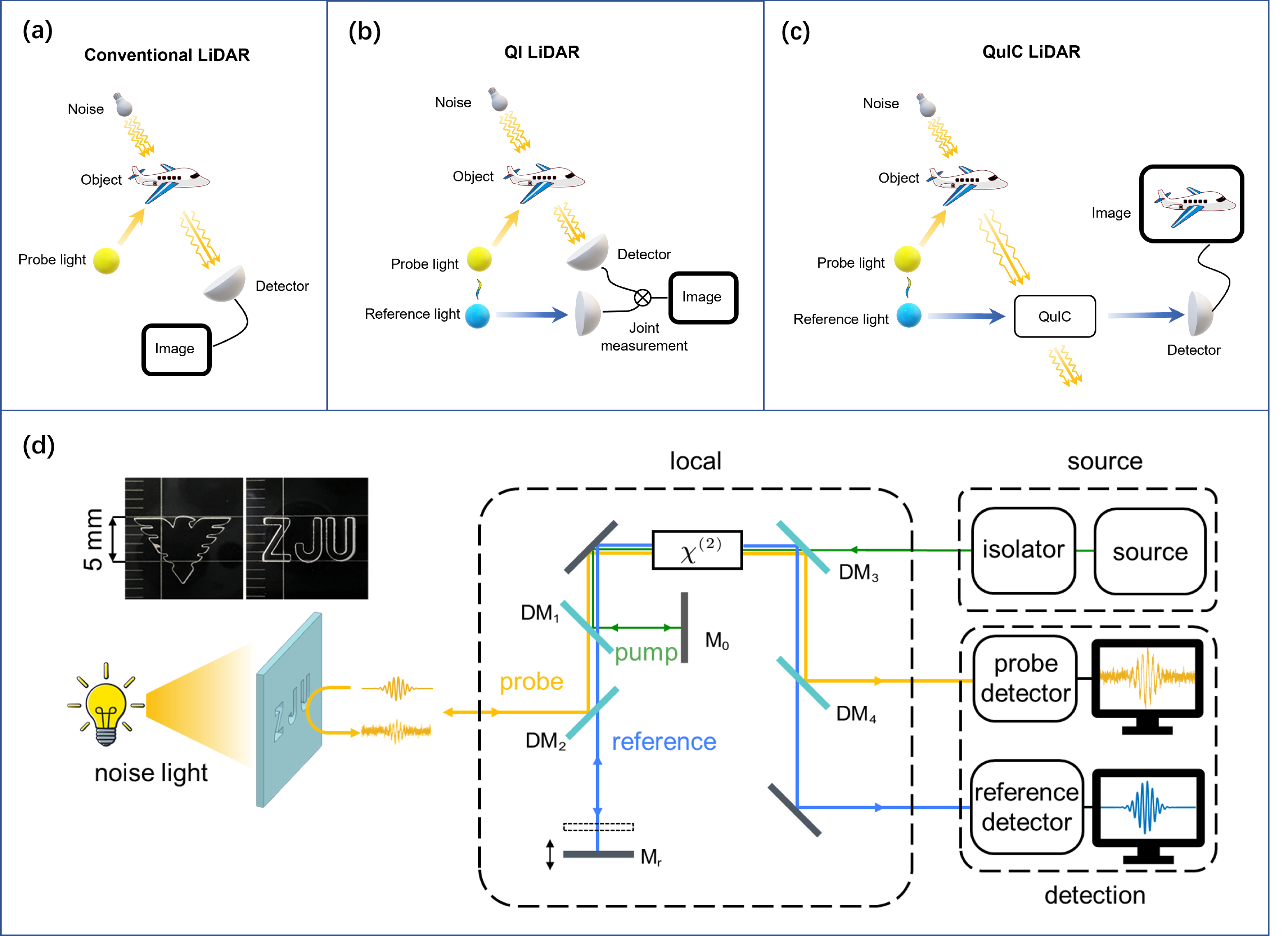}
\end{center}
\caption{
{Physical principles for conventional, QI and QuIC LiDARs.} (a) Conventional-illumination-based LiDAR. (b) Quantum illumination LiDAR, where the SNR is enhanced by joint measurement. (c) QuIC LiDAR, instead of the probe light which is mixed with noise, a locally generated reference light, which
benefits from quantum induced coherence, is detected. (d) The setup consists of three modules (enclosed in dashed lines), the pump laser, the local scanning module and the detection module. A collimated 532 nm laser (red line) is used to pump a PPLN crystal to generate SPDC entangled photon pairs whose frequencies can be tuned by adjusting the temperature of the crystal. Here the wavelengths of the reference (blue line) and probe (yellow line) lights are centered at 893 and 1316 nm. The two light beams are separated by a dichroic mirror DM$_{2}$. The probe beam is sent to the object while the reference beam is sent to a scanning mirror M$_r$. The reflected probe, reference and pump beams go through the crystal a second time. We detect the reference beam for QuIC LiDAR with a Si-based CMOS camera. The probe light is also detected with an InGaAs detector to compare the performance of QuIC LiDAR and traditional OCT.
}
\label{fig1}
\end{figure*}

Quantum properties of light such as entanglement and squeezing can be used to enhance the sensitivity and signal-to-noise ratio (SNR) in light detection and ranging (LiDAR) \cite{Vittorio_2001, giovannetti_2002,lloyd_2008,maccone_2020}. Such a quantum LiDAR can be accomplished by using quantum illumination (QI) of entangled photons combined with correlation detection \cite{lloyd_2008, maccone_2020,tan2008quantum, shapiro2020quantum, nair2020fundamental,lanzagorta2011quantum,wilde2017gaussian,sanz2017quantum,zhang2013entanglement,Zhuang2022}, where one beam of entangled photons from spontaneous parametric down conversion (SPDC) serves as a reference while the other beam is sent to probe the presence of an object. The extra information provided by the reference photons in correlation detection significantly improves the SNR in detecting the object in noisy and lossy environment \cite{lopaeva2013experimental,pirandola2018advances,zhang2015entanglement,zhuang2017optimum} (see Fig.~\ref{fig1} (a) and (b)). The spatial and temporal correlation of entangled photons can be used to extract the image \cite{gregory2020imaging} and distance \cite{zhao2022light} of the object based on joint measurement, which has been experimentally extended to the microwave regime \cite{barzanjeh2020microwave, chang2019quantum,barzanjeh2020microwave}. 
However, in the above described approach the noise from the object still enters the detector, which can be jammed by saturation attacks when the noise is much stronger than the signal. 

Quantum induced coherence, first demonstrated in the Zou-Wang-Mandel (ZWM) experiment \cite{zou1991induced} without employing coincidence detection, has been used to 
image objects with undetected photons \cite{lemos2014quantum,paterova2020hyperspectral,kviatkovsky2020microscopy,paterova2020quantum,hochrainer2022quantum,hochrainer2017interference}. In such a setup entangled photons are generated in the SPDC of two nonlinear crystals. 
The information on the origin of the entangled photon pairs plays a key role in the single-photon interference of the SPDC light. 
The interference of signal photons depends on whether we in principle know which crystal generates the photon pairs. By overlapping the idler modes of the two crystals, which-way information of the signal photons is erased \cite{zajonc1991quantum,scully1982quantum,scully1991quantum}, the single-photon interference of the signal photons can be observed. On the other hand, if an object is placed between the two crystals to block the idler photons, we can identify which crystal generates the entangled photon pairs, such that the single-photon interference pattern of the signal photons disappears. This mechanism has also been widely used in spectroscopy \cite{kutas2021quantum,kalashnikov2016infrared,paterova2020nonlinear}, optical coherence tomography \cite{paterova2018tunable,vanselow2020frequency,mukamel2020roadmap,gilaberte2021video} and holography \cite{topfer2022quantum}. In these applications visible photons are usually used to trigger the detector while their entangled infrared partners are used to interact with the object, such that infrared imaging and spectroscopy are achieved with visible light detectors.

In this Letter, we design and implement a quantum induced coherence (QuIC) LiDAR based on the ZWM interference. It has a substantial advantage with respect to previous QI LiDAR proposals since QuIC LiDAR detects the image and distance of objects from the visibility of single-photon interference fringes rather than two-photon coincidence counting, significantly enhancing the detection efficiency, shortening the collection time and improving the ranging resolution (see Fig.\ref{fig1} (b) and (c)).  More importantly, since we detect the reference photons rather than the probe photons that are used to interact with the object, such a scheme is intrinsically immune to background noise and intentional jamming. 
We demonstrate the ranging and imaging ability of QuIC LiDAR and show its robustness in LED and laser backgrouond noise. {This proof-of-principle demonstration obtains a 20 dB enhancement in noise resilience of the QuIC LiDAR, in comparison with that using classical coherence.} QuIC LiDAR can be readily extended to mid-infrared and terahertz regime, where signal detection and background noise are challenging problems.


\begin{figure*}
\begin{center}
\includegraphics[width=0.7\textwidth]{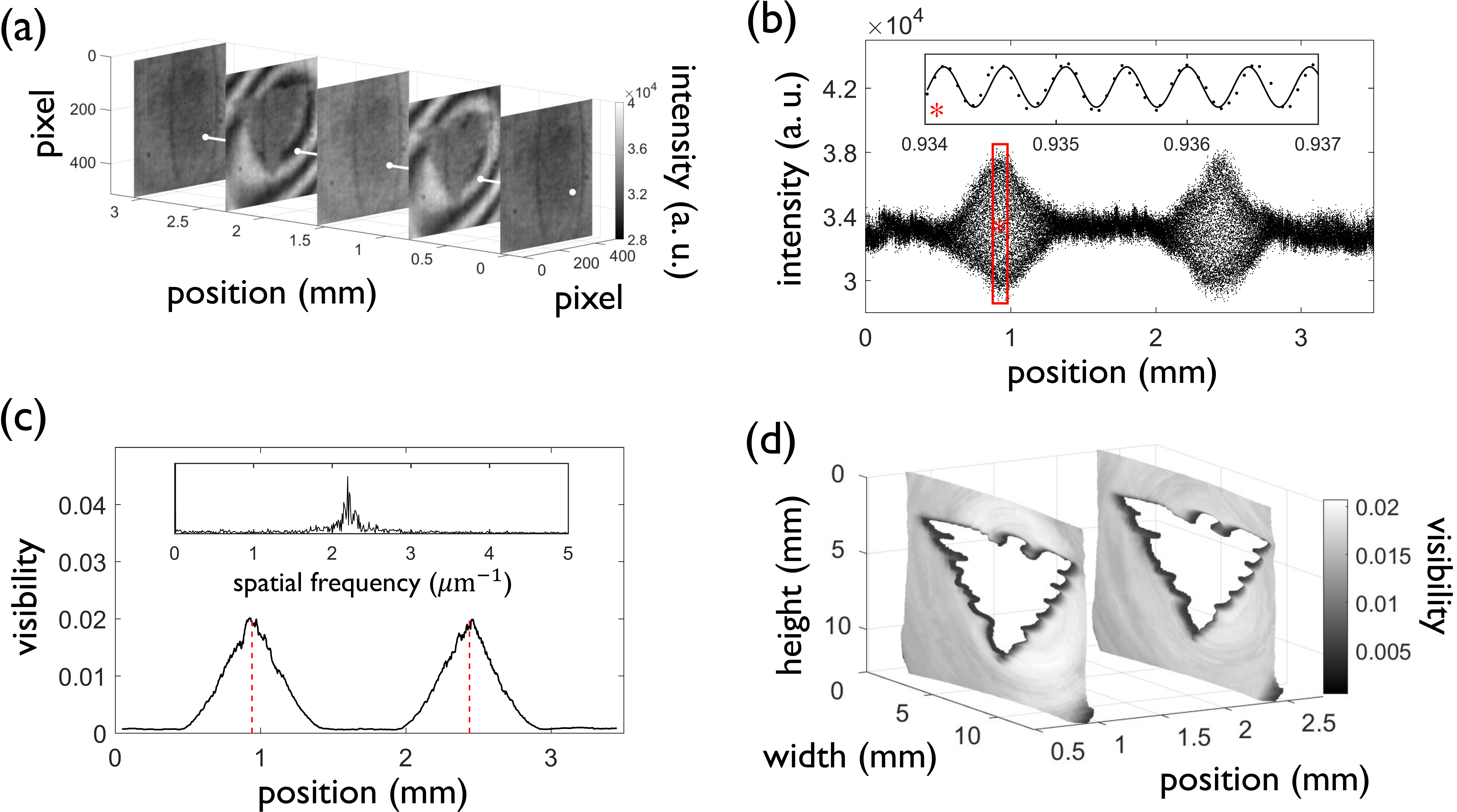}
\end{center}
\caption{
Position acquisition of QuIC LiDAR. 
(a) Real-time images captured by the CMOS camera during the scanning of M$_r$. The two images carrying interference patterns correspond to the two surfaces of the plate. The exposure time of each image is 50 ms. The scanning step is 60 nm with a total scanning length 3.5 mm. 
(b) Interference fringes along the white lines in (a). The inset shows the interference near the position marked by the asterisk. 
(c) Interference visibility obtained from STFT with an integration window  100 $\mu$m. The step in the horizontal axis is 1 $\mu$m. Inset: the Fourier transform of the interference in the red frame in (b). The maximum peak value in the spatial frequency range from 2.0 to 2.4 $\mu$m$^{-1}$ is extracted as the visibility. The two peaks in the main frame with the weights being marked by red dashed lines indicate the positions of the two surfaces. (d) Images on the two surfaces obtained from QuIC LiDAR.
}
\label{fig2}
\end{figure*}

\begin{figure*}[t]
\begin{center}
\includegraphics[width=0.8\textwidth]{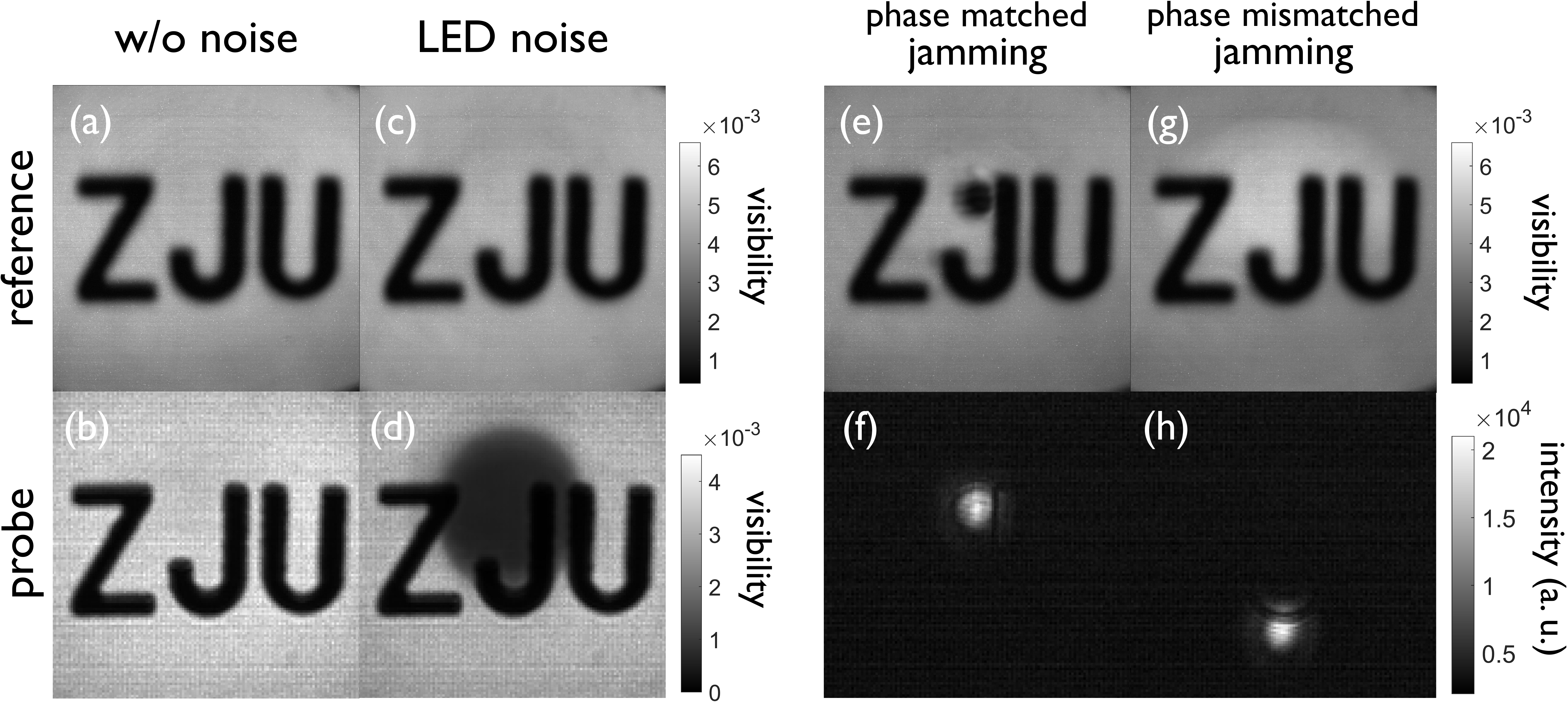}
\end{center}
\caption{
Noise robustness of QuIC LiDAR.  The images in the upper (lower) row are extracted from the reference (probe) light cameras. (a) and (b) are the images without background noise. (c) and (d) are the images with background LED light. (e), (f), (g) and (h) are images with a jamming laser shining on the sample. In (e) and (f) the direction of the laser satisfies the phase matching condition for stimulated PDC in the PPLN crystal, while in (g) and (h) there is a phase mismatch. Except for a phase matched jamming attack with a laser source, the QuIC LiDAR images from the reference light are unaffected by the noises. In comparison, the probe light images are prone to all the noises. 
The powers of the reflected probe beam, the LED light and the laser before entering the probe camera are 72 nW, 44 nW and 2.6 $\mu$W. The exposure time is 6 ms for (a), (c), (e) and (g), 15ms for (b) and (d), and 20 $\mu$s for (f) and (h).  Here (f) and (h) are unprocessed images from the probe camera.
{The reference image has lower resolution compared with the one from the probe beam, due to the imperfection of the transverse phase matching in the crystal, which is restricted by the crystal aperture and the pump beam waist size \cite{fuenzalida2022resolution}.}
}
\label{fig3}
\end{figure*}

Three light beams are used to build the QuIC LiDAR. A pump laser beam pumps a nonlinear crystal, which generates a probe (idler) beam and a reference (signal) beam in SPDC. Both the pump and reference beams are kept as local while the probe beam is sent out to detect the object. The probe beam reflected from the object is collected to go through the nonlinear crystal a second time together with the locally reflected pump and reference beams in a Michalson configuration \cite{chekhova2016nonlinear} of the ZWM experiment \cite{zou1991induced} (see Fig.~\ref{fig1}(d)). We use a single-frequency 532 nm (10 kHz linewidth) laser to pump a type-0 periodically poled lithium niobate (PPLN) crystal (length 2 cm). The narrow linewidth (long coherence length) of the pump light guarantees that the coherence in the two SPDC processes is maintained. 
We observe interference when the light paths of the reference and probe beams (from leaving to re-entering the nonlinear crystal) are equal within the coherence length of the SPDC modes. 
The light intensity of the reference beam at the detector (see Supplementary Materials),
\begin{equation}
\begin{aligned}
I_r \propto \eta^2 [1 +\gamma(\tau)|r_p| \cos(\phi_p+\phi_r-\phi_0)],
\end{aligned}
\label{signal}
\end{equation}
where $\eta$ is the probability amplitude of generating an entangled photon pair in a single pass through the nonlinear crystal, $r_p$ is the reflectivity amplitude of the probe beam from the object, $\phi_p$, $\phi_r$ and $\phi_0$ are the accumulated phases of the probe, reference and pump beams in the round loops from leaving to returning to the PPLN crystal. Here we assume $\gamma(\tau)=\exp(-\tau^2/2\sigma^2)$ be the overlapping function between the returned probe and reference light modes  with $\tau$ being their traveling time difference. The interference fringes are significant only when $\tau$ is smaller than $1/\Delta\omega$, where $\Delta\omega$ is the frequency bandwidth of the SPDC light. We obtain the distance of the object by scanning the light path of the reference beam to achieve maxima in the interference visibility,
\begin{equation}
    V=|r_p| \gamma(\tau).
\end{equation}
Compared to QI LiDAR which relies on coincidence counting with signals proportional to $|r_p|^2$, the visibility in QuIC LiDAR is proportional to $|r_p|$, i.e., linear with the reflectivity amplitude of the object \cite{wang1991induced}. Such a linear scaling makes QuIC LiDAR highly advantageous for detecting surfaces with low reflectivity. This sensitivity enhancement is the result of the mixing between the returned probe light and the local reference light before being detected, similar to the mixing between the signal and the local oscillator in homodyne detection. 


The performance of QuIC LiDAR is tested with two silica plates with hollowed symbols (Fig~\ref{fig1} (d)). We scan the reference light path until observing interference patterns on the CMOS camera, which indicates that an object reflects the probe light from a distance the same as the one between M$_r$ and the PPLN crystal. We find two such positions corresponding to the two surfaces of the silica plates (see Fig.~\ref{fig2} (a) and (b)). The envelopes of the interference fringes during the scanning have a spatial width 0.4 mm, which is determined by the coherence length of the entangled photons and sets a limit on the ranging resolution of QuIC LiDAR. This is in the same spirit of improving resolution through mode gating \cite{Ansari2021}, where the gating mode is provided by the reflected probe photons. In order to filter the noise in determining the distance of the object, we perform short-time Fourier transform (STFT) with a scanning window 100 $\mu$m and obtain the visibility as a function of the scanning distance (Fig.~\ref{fig2} (c)). 
The weights of the two peaks are regarded as the expected distances of the two surfaces, such that a 3D reconstruction of objects is achieved with a lateral resolution 412 $\mu$m and a ranging accuracy 5.1 $\mu$m (see Fig.~\ref{fig2} (d) and Supplementary Materials), which can be considered as a quantum version of the optical coherence tomography (OCT) \cite{paterova2018tunable,Valles2018,vanselow2020frequency}. The ranging accuracy is higher than that of the LiDAR relying on the measurement of time of flight (including QI LiDAR), which is limited by the jitter time (tens of picoseconds resulting in milimeter uncertainty) of electronic devices. The visibility of the interference patterns varies across the beam due to the finite lateral coherence of the SPDC light.  

While QI can significantly enhance the SNR when the noise is much lower than the signal \cite{lloyd_2008,lopaeva2013experimental}, its performance in LiDAR quickly deteriorates when the background noise increases to the same level of the signal. It is also prone to jamming attack since all photons from the object are detected (see Fig.\ref{fig1}(b)). In comparison, QuIC LiDAR is superior in noise resilience, originating from the essence of ZWM experiment, i.e., the position of the object is obtained by detecting the local reference light instead of the reflected probe light (see Fig.\ref{fig1}(c)). The ambient noise that has the same wavelength as the probe light, although can go into the local module, is filtered by the dichroic mirror DM$_4$ before the reference light detector. Such noise neither has any effect on the interference pattern of the reference light, since they cannot alter its which-way-information. On the other hand, the noise that has the same polarization and wavelength as the detected reference light is filtered by another dichiroic mirror, DM$_2$, before entering the local module.

We test such noise resilience of QuIC LiDAR by shining the sample from behind with a LED light and a laser. In order to make a comparison, we also extract the images from the probe light following the same procedure as that of the reference light to mimick the performance of ordinary optical coherence tomography. Without background noise, the two images extracted from the reference and probe light cameras are similar except for a slight difference in resolution \cite{fuenzalida2022resolution} (Fig.~\ref{fig3}(a) and (b)). We use LED light with a similar wavelength as that of the probe light to simulate background noise. The spectrum power density of the LED is about 7 mW/m$^2$/nm, much lower than that of the sun (300 mW/m$^2$/nm). As a result, the noise enters the probe light camera and substantially obscures the image (Fig.~\ref{fig3}(d)), while no difference is detected by the reference light camera (Fig.~\ref{fig3}(c)), which demonstrates the robustness of QuIC LiDAR against the background noise. 

\begin{figure}[t]
\begin{center}
\includegraphics[scale=1,angle=0,width=0.48\textwidth]{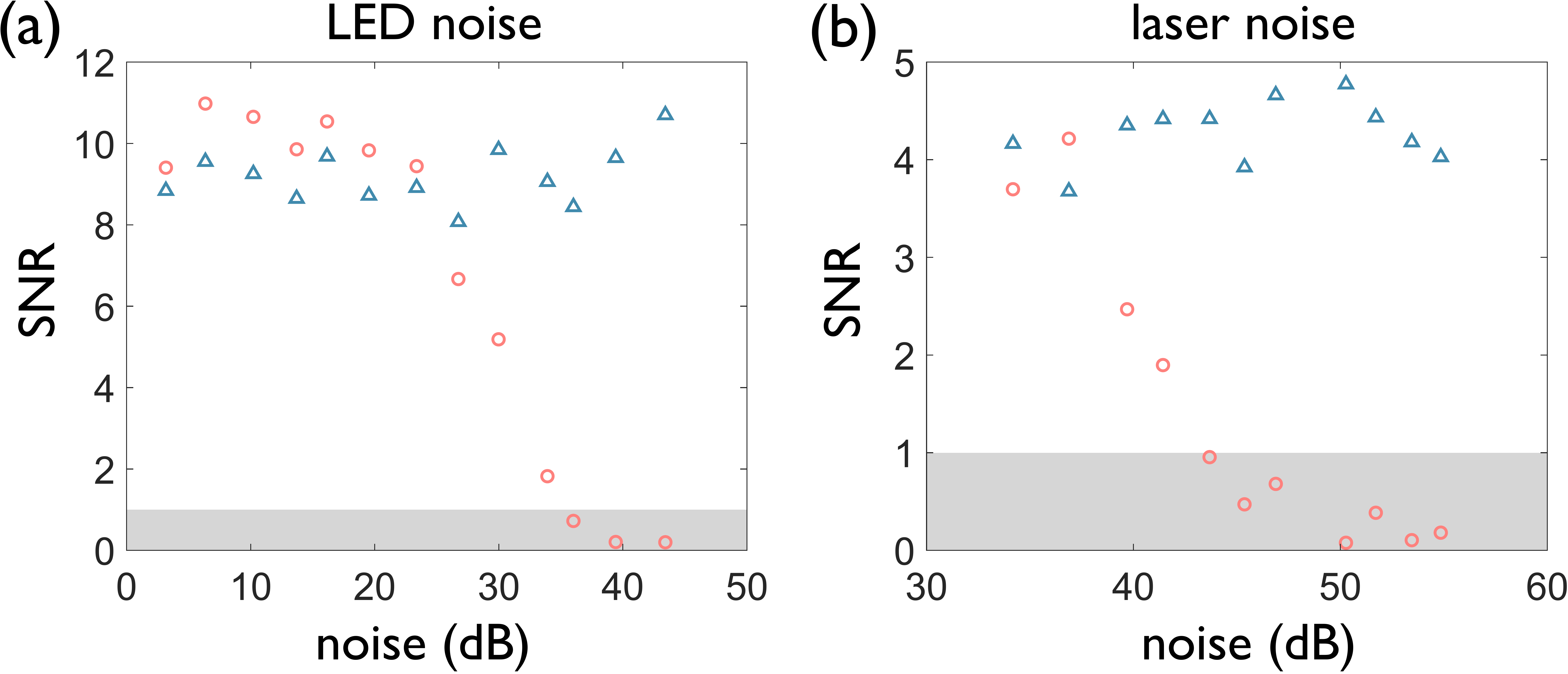}
\end{center}
\caption{
{The SNR of QuIC LiDAR in LED (a) and laser (b) noises. In (a), we compare the SNR in QuIC LiDAR (blue triangles) and in direct detection of the probe light (red circles). For noise level below 23.4 dB, the LED noise is weaker than the systematic noise,
such that the SNR of the probe light remains unchanged. For higher LED noise, the SNR of the probe light decreases, while the one of the QuIC LiDAR remains unchanged. In (b), we simulate saturation jamming attacks by sending a laser back to the crystal, which induces stimulated PDC noise. The red circles and blue triangles are the SNRs of the reference light in the areas satisfying and violating the phasing matching condition (of the jamming laser), respectively. The powers of probe and noise light are measured by an InGaAs power meter with an iris covering the region of interest on the camera.} 
}
\label{fig4}
\end{figure}

The only practical way to jam QuIC LiDAR is to send a laser that has the same frequency and polarization as the probe light back to the local module. The stimulated parametric down conversion (PDC) light can jam the detector of the reference light (see the black spot in Fig.~\ref{fig3} (e)). However, {in order to jam the LiDAR through the PDC process, the pump and jamming lasers have to overlap in space, be colinear, and the wavelength and propagating direction of the jamming laser should satisfy the phase matching conditions for a given crystal temperature.} A slight deviation of the laser from the phase matching angle substantially reduces the stimulated PDC noise. As a result, the jamming is only efficient in specific directions covering a few pixels, leaving other pixels undisturbed (see Fig.~\ref{fig3} (e)). Such a strict requirement on phase-matching makes it difficult to jam QuIC LiDAR, and also allows us to counter fight such an attack by adjusting the orientation or the temperature of the PPLN crystal (Fig.~\ref{fig3} (g)). In comparison, the probe light camera can be easily jammed by the laser (see Fig.~\ref{fig3} (f) and (h)).

To quantitatively characterize the noise robustness of QuIC LiDAR, we measure the SNR, $\mathcal{F}_s / \mathcal{F}_n $, where $ \mathcal{F}_s $ and $ \mathcal{F}_n $ are the signal peak and noise baseline in the Fourier transform spectrum (see Fig.S4 in the Supplementary Materials), at different noise levels defined in decibel, $ 10 {\rm log}_{10} (  P_n / P_p ) $, where $ P_n $ and  $ P_p $ are the powers of the noise and reflected probe light, respectively (see Fig.\ref{fig4}). We compare the SNR of the interference of the probe and reference light to quantify the advantage of QuIC LiDAR in noise resilience. For the LED noise, the SNR of the probe light starts dropping at a noise level 23.4 dB until going below the detection line at 36 dB, while the SNR of the reference light is unaffected (see Fig.~\ref{fig4}(a)). While only 43.3 dB of noise power can be imposed with our LED, QuIC LiDAR can resist much stronger noise. For saturation attacks from a laser source, the stimulated PDC light at the phase matching direction can reduce the SNR of the reference light to 1 with a laser noise 43.7 dB. However, in pixels where the phase matching condition cannot be satisfied, the SNR remains unaffected, as shown by the blue triangles in Fig.~\ref{fig4}(b).

In conclusion, we demonstrate that quantum induced coherence can be used to build a quantum LiDAR that has superior noise resilience compared with QI LiDAR which relies on coincidence counting. Although a similar setup has been used in imaging, spectroscopy and optical coherence tomography, featured by visible light detection with infrared light probing, its ability in ranging and battling noise remains largely unnoticed. {With the freedom in selecting wavelengths in SPDC \cite{hochrainer2022quantum}, we can build mid-IR and terahertz QuIC LiDAR by using proper crystals (e.g., AgGaS$_2$ for mid-infrared up to 13um) \cite{kumar2021mid, kutas2021quantum,mukai2022quantum,paterova2022broadband}. In such wavelengths the LiDAR detection is better concealed and difficult to jam because efficient detectors and lasers are rare. QuIC LiDAR can be used to meet the challenges of background noise and efficient detection in those wavelengths. For applications in microwave regime where background noise brings a huge challenge, we can integrate the microwave-to-optics
conversion \cite{chang2019quantum, andrews2014bidirectional, tu2022high} to the current design or use microwave-optical entangled light sources \cite{Rueda2019}. The weak intensity of SPDC light makes it difficult to detect objects with low reflection. A solution to this problem is to use an SU(1,1) interferometer by adding a seed to our current setup to improve the signal-to-noise ratio \cite{heuer2015complementarity}. The QuIC LiDAR described here can be integrated with on-chip entangled photon sources and delay lines \cite{ono2019observation} to make it small and compact to be used in modern transportation vehicles. 
}

This work was supported by the National Natural Science Foundation of China (Grants No.~11934011), Zhejiang Province Key Research and Development Program (Grant No.~2020C01019), the Strategic Priority Research Program of Chinese Academy of Sciences (Grant No.~XDB28000000), and the Fundamental Research Funds for the Central Universities. VVY acknowledges partial support from the NSF (CMMI-1826078), AFOSR (FA9550-20-1-0366), and NIH (1R01GM127696, 1R21GM142107, 1R21CA269099).


$\ $

$^{*}$xuxingqi@zju.edu.cn

$^{\dagger}$dwwang@zju.edu.cn

\bibliography{apssamp}

\end{document}